# Recovery of parameters of fast nonlocal heat transport in magnetic fusion plasmas: testing a model of waves with high internal reflections

A.B. Kukushkin[1,2,3], P.A. Sdvizhenskii[1], A.V. Sokolov[4], P.V. Minashin[1]

[1]National Research Center 'Kurchatov Institute', Moscow, 123182, Russian Federation
[2]National Research Nuclear University MEPhI, Moscow, 115409, Russian Federation
[3]Moscow Institute of Physics and Technology, Dolgoprudny, Moscow Region, 141700, Russian Federation
[4]Institute for Information Transmission Problems (Kharkevich Institute) of Russian Academy of Science, 127051 Moscow, Russian Federation

## Abstract

Our analysis of the model [J. Phys.: Conf. Ser. **941** (2017) 012008] elaborated for interpreting the initial stage of the fast nonlocal transport events, which exhibit immediate response, in the heat diffusion time scale, of the spatial profile of electron temperature to its local perturbation, shows that the nonlocal transport by electromagnetic (EM) waves needs too high reflectivity of vacuum vessel walls to describe the experimental data. Here we try another model, which assumes high internal reflections and is compatible with the "wild cable" transport of TEM waves along magnetically-bound skeletal nanostructures. An inverse problem for recovery of the source and sink of waves, and internal reflectivity, is formulated and solved. Preliminary results of analyzing the data from tokamaks JET and TFTR, and stellarator LHD are presented.

## 1. Introduction

To describe the anomalous heat transport in the magnetized thermonuclear plasma the superdiffusion formalism was suggested (namely, the equation, integral in space coordinates, with a long-tailed kernel and, respectively, dominance of carriers with a long free path). It includes steady-state transport of heat by electron/ion Bernstein [1] and electron cyclotron waves [2]; transport of temperature perturbation by electron Bernstein waves [3], which was described by Biberman-Holstein equation known from the theory of atomic excitation's transport in spectral lines of atoms/ions; model of fractional derivatives for heat perturbation transport (see references in the reviews [4-6]). One of the most interesting examples of fast nonlocal transport is the unusual direction of a heat flux, namely, the instantaneous (in the heat diffusion time scale) temperature increase in the plasma center in case of fast cooling the plasma periphery (so called "cold pulse" experiments in tokamaks and stellarators, [4-6]) or an inverse process — the instantaneous temperature fall in the plasma center in case of fast heating the plasma periphery [7].

An inverse problem of recovering the integral equation kernel was formulated, extending/modifying the works [8 — 10], and solved in [11, 12]. The equations [11] described the nonlocal (superdiffusive) energy transfer caused by emission and absorption of electromagnetic (EM) waves with a long free path and strong reflection of waves from the walls of vacuum chamber. The developed model was used for interpretation of the "cold pulse" experiments in stellarator LHD [13, 14] and tokamak TFTR [4, 15].

Here we modify the model [11, 12]. Our analysis of this model, based on the approach [8, 16-18] to modeling the nonlocal heat transport by the long-free-path electromagnetic (EM) waves (the results will be reported elsewhere), shows that the nonlocal transport by the EM waves needs too high reflectivity of vacuum vessel walls to describe the experimental data. Also, cooling of the periphery almost the same way in similar (not identical though) discharges not necessarily produces the heating of the core, an instant cooling of the core is also possible (cf. e.g. the data from tokamak JET experiments, Fig. 23 in [19]). Therefore, a modification of the model [11, 12] is needed to allow for the both above-mentioned issues and describe the most common feature of the observed phenomena of fast nonlocal transport, namely, the instant response, in the heat diffusion time scale, and its possible dependence on the parameters not identified yet in the existing transport models.

Here we suggest and analyze a model, which assumes high internal reflections of the waves and is compatible with the "wild cable" transport of TEM waves along magnetically-bound skeletal nanostructures [20, 21] (Sec. 2). An inverse problem for recovery of the source and sink of waves, and internal reflectivity, is formulated (Sec. 3) and solved for particular experimental data (Sec. 4). Preliminary results of analyzing the data from tokamaks JET [19] and TFTR [4, 15], and stellarator LHD [13, 14] are presented.

## 2. A model of nonlocal heat transport by EM waves with high internal reflections

A model of nonlocal heat transport by the EM waves with high internal reflections is compatible with the "wild cable" transport of transverse EM (TEM) waves along magnetically-bound skeletal nanostructures [20, 21]. This model differs from the model [11, 12] in the following points.
  A. The reflection of waves takes place inside plasma that is described by the high reflectivity from the plasma boundary; also the internal reflections inside the wild cable network (e.g. those from the "pedestal" in toroidal plasmas and from other internal transport barriers) produce substantial slowing of the waves with respect to the speed of light in vacuum despite the TEM waves travel in vacuum channels around skeletal structure in "wild cables" (vacuum channels are sustained by the internal and external sources of EM field in plasma; vacuum channels protect skeletal structure from high-temperature plasma by the Miller ponderomotive force).
  B. The volume-average intensity of waves is again assumed to be homogeneous and isotropic; however, the local intensity on the outer boundary of the wild cable should be equal to local plasma pressure.
  C. The sink function of the energy carriers (TEM waves) is known and described by the absorption of waves by the plasma on the outer boundary of the wild cable (i.e. the confinement of the waves in vacuum channel is accompanied by the loss of wave energy due to finite electric conductivity of plasma, the high-frequency surface impedance of the plasma around vacuum channel). Therefore, the source function is retained to be a variable in the inverse problem while the uncertainty of the sink function is limited to the uncertainty of the parameters of the wild cable network (the latter is described by the sought-for function $X(\rho, t)$ which depends on the spatial structure of the wild cable network and on the group velocity of waves inside the network with account of internal reflections).

Under these assumptions, we arrive at the following energy balance equation for describing the initial stage of nonlocal heat transport events in the core plasma after a fast local perturbation of the plasma periphery at a time moment $t_0$:

$$\left[A_{bs}(\rho, t)\sqrt{I(t)} - A_{bs}(\rho, t_0)\sqrt{I(t_0)}\right] = \Pi(\rho, t),$$

$$0 < \rho < \rho_{max}(\sim 0.5) < \rho_p(\sim 0.9 - 0.95), \qquad t_0 < t_{min} \leq t \leq t_{max} \tag{1}$$

$$A_{bs}(\rho,t) = 0.2\sqrt{a(m)}\sqrt{\frac{\Lambda}{10}\left\{\frac{Z(\rho,t)\,n(\rho,t)}{[T(\rho,t)]^{1/2}}A(\rho,t)X(\rho,t)\right\}^{1/2}},$$

(2)

where $\rho$ is the normalized minor radius coordinate in the 1D representation of the energy density balance; $T$, electron temperature in keV; $n$, electron density in $10^{19}$ m$^{-3}$; Z, ion's average charge, a(m), effective minor radius of plasma in meters; $\Lambda$, Coulomb logarithm; and the contribution of total plasma pressure is described by the function

$$A(\rho,t) = \left[1 + \frac{T_i(\rho,t)}{Z(\rho,t)T(\rho,t)}\right],$$

(3)

where the subscript *i* stands for ions. For the normalized intensity *I* of waves, we have a simple equation

$$\sqrt{I(t)} = \frac{\sqrt{4\,\bar{q}(t)F_{out}(t) + \left(\bar{A}_{bs}(t)\right)^2} - \bar{A}_{bs}(t)}{2\,F_{out}(t)},$$

(4)

where $\bar{q}(t)$ is the volume-average source of waves; $F_{out} = 2\pi(1-R)$ describes the transmission coefficient (R is the reflection coefficient), the normalized volume-average absorption function $A_{bs}$ describes the sink of waves:

$$\bar{A}_{bs}(t) = \int_0^{\rho_p} A_{bs}(\rho,t)\,\frac{2\rho d\rho}{\rho_p^2}.$$

(5)

The above nonlocal transport model is applicable for the mean free path of waves (with respect to absorption) not less than the effective minor radius of the plasma, that gives the restriction

$$0 < \frac{\bar{A}_{bs}(t)}{\sqrt{I(t)}} < 1$$

(6)

Also, the limits for $F_{out}$ are as follows:

$$0 < F_{out}(t) < (F_{out})_{max} = 2\pi,$$

(7)

The right-hand side of Eq. (1) assumes that at initial stage only the electron energy balance is important, and the electron density in the core is not disturbed (cf. e.g. [4]):

$$\Pi(\rho,t) = \frac{3}{2}n(\rho)\frac{\partial}{\partial t}[T(\rho,t)]$$

(8)

If the density is measured in $10^{19}$ m$^{-3}$, and temperature in keV, the functions $\Pi$ and $\bar{q}$ are measured in 1.6 kW/m$^3$.

For the volume-average source, we introduce the limitation

$$0 < \bar{q}(t) < f \frac{Q_{tot}}{V_{plasma}}, \quad f < \sim 0.1 - 0.2,$$

(9)

where $Q_{tot}$ is the total power of plasma heating, and $V_{plasma}$ is the total plasma volume.

Analysis of available experimental data and a high degree of freedom for the sought-for parameters suggests that at the present stage of analyzing the above model, it's appropriate to start with a simplified version of Eq. (1) via neglecting the time dependence of the absorption function. This gives the following equation:

$$A_{bs}(\rho)\left[\sqrt{I(t)} - \sqrt{I(t_0)}\right] = \Pi(\rho, t).$$

(10)

Also, we neglect the time dependence of the source function $\bar{q}(t)$.

## 3. Inverse problem and the method of solving

The inverse problem is formulated as follows.
- Input parameters: $n(\rho, t), T(\rho, t)$.
- Output (sought-for) parameters: $A_{bs}(\rho), F_{out}(t)$, including $F_{out}(t_0)$, and $\bar{q}$.

Additionally, the variation of $F_{out}(t)$ is allowed within the range of one order of magnitude, and the ratio of the wave power absorbed in the core and in the remaining plasma is less than a factor (taken here equal to 2).

Numerical calculations are carried out by the method of sequential correction and identification of mathematical model parameters in Eqs. (1) — (10). A method based on a sequential, regularized approximation of the input data by parametric models is used. This approach is ideologically close to the methods of Tikhonov's regularization [22] (in this paper there is also regularization by smoothness). The regularization criterion is the smoothness of the approximating functions. Smoothness is treated as one of the important characteristics of the solution quality. The general scheme for estimating the modeling error is the cross-validation. Similar approaches to data processing can be found in [23]. The general scheme of the method proposed by A.V. Sokolov [24] has already been successfully applied to various problems, including the processing of geophysical data [25], time dependence of $CO_2$ flux [26], simulation of pine transpiration [27].

To estimate the quality of parametric models and to choose the direction of their improvement, a large number of auxiliary subtasks has to be solved. Each step of the proposed method is a two-level finite-dimensional non-linear problem of mathematical programming. At the "lower" level, it is necessary to solve a large set of independent subtasks, and this phase of calculations can be significantly accelerated when computing is transferred to a distributed computing environment. To do this, the system of services for solving the non-linear problems of mathematical programming on the Everest platform was used [28], where the package pool (Ipopt, https://projects.coin-or.org/Ipopt) was deployed to solve such problems. For the software implementation of the method, including the description of tasks and the data exchange with packages, the Python language and the Pyomo optimization modeling package are used [29, 30]. The general scheme of the Pyomo application in the Everest system of services coincides with that already used for integration of the optimization modeling language AMPL [31] into the Everest. A more detailed information on the proposed numerical method can be found in [32], [24]. The balanced identification method we use here is described in [27].

## 4. The results of data analysis for JET, TFTR, and LHD

The simplification of Eq. (1) to the form of a separated temporal and spatial dependencies of the left-hand side imposes the limits on the accuracy of obeying the Eq. (10). On the other hand, one can obtain relatively simple and reliable estimates of possible physically-meaningful solutions of Eq. (1).

We show the results for minimal value of volume-average source where the accuracy (relative mean square deviation (MSD) of the left-hand side from the right-hand side in Eq. (10) in the core plasma at the chosen interval of time after perturbation of the periphery) is not worse than the best accuracy of fitting the right-hand side of Eq. (10) with the product of arbitrary function of space coordinate and arbitrary function of time.

### 4.1. Tokamak JET

The data of Fig. 23 from [19] are processed in the frame of the model (1)-(10). The data from subsequent discharges show that under almost the same conditions (the power of auxiliary heating Q differs by a factor of about 2, the pellet injection is similar) the plasma dynamics exhibits quite opposite trends of temperature evolution in the central, core plasma: an increase in the shot #53487 and a decrease in the shot #53488. The behavior of the edge plasma is different in these shots so that one can assume different reflectivity of waves from the plasma edge. The inverse problem results give a quantified picture of the expected role of the transmission coefficient and evaluate the profile of absorbed power and the dependence of the accuracy on the value of volume-average source.

For shot #53487, time interval from 7.0 s to 7.05 s is taken. The solution of the inverse problem gives $\bar{q} = 5$ ($f \sim 0.2$), and relative MSD = 34.1 %. The results for sink and transmission functions are shown in Figure 1.

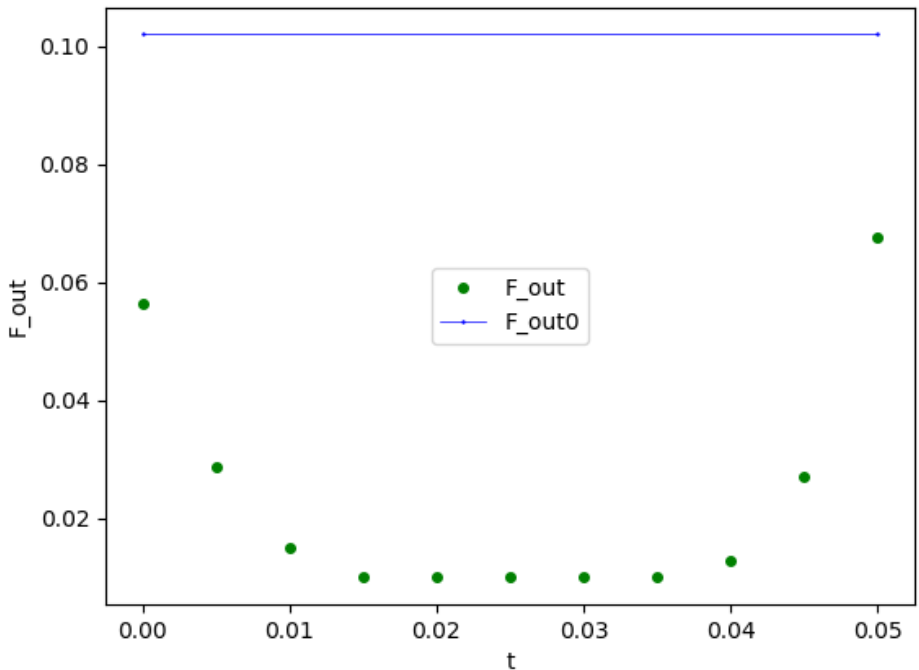

Fig. 1(a). Time dependence of transmission coefficient $F_{out} = 2\pi(1-R)$. The value before perturbation is also shown.

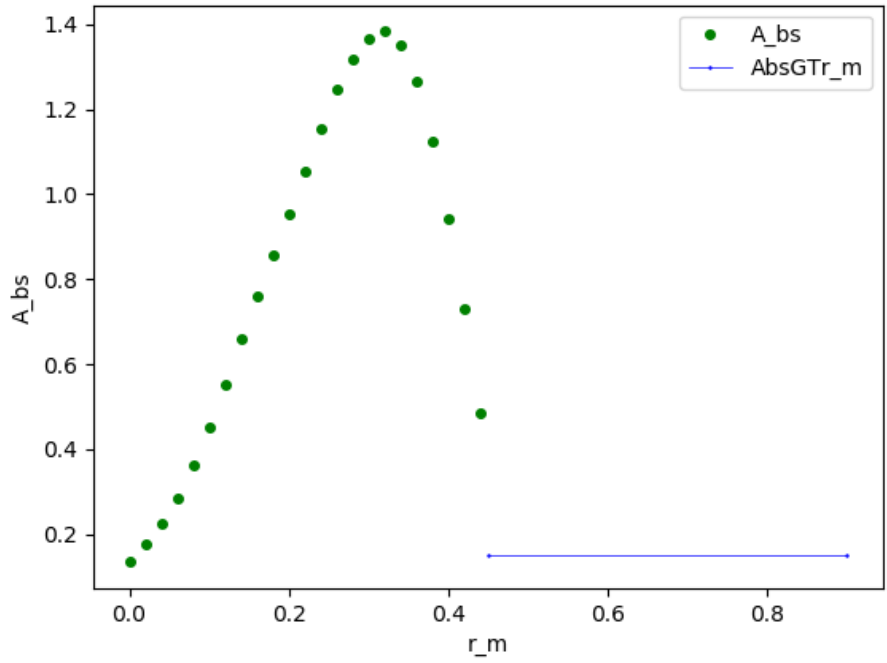

Fig. 1(b). The profile of absorption power as a function of the effective minor radius coordinate in the core plasma. The volume-average value in the residual plasma is shown.

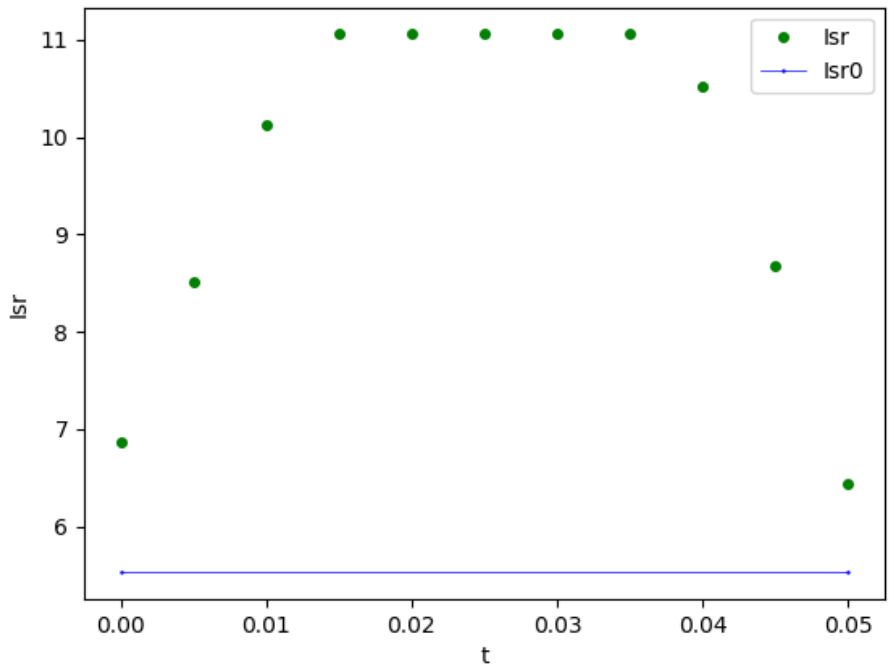

Fig. 1(c). Time dependence of the square root of normalized intensity, Eq. (4).

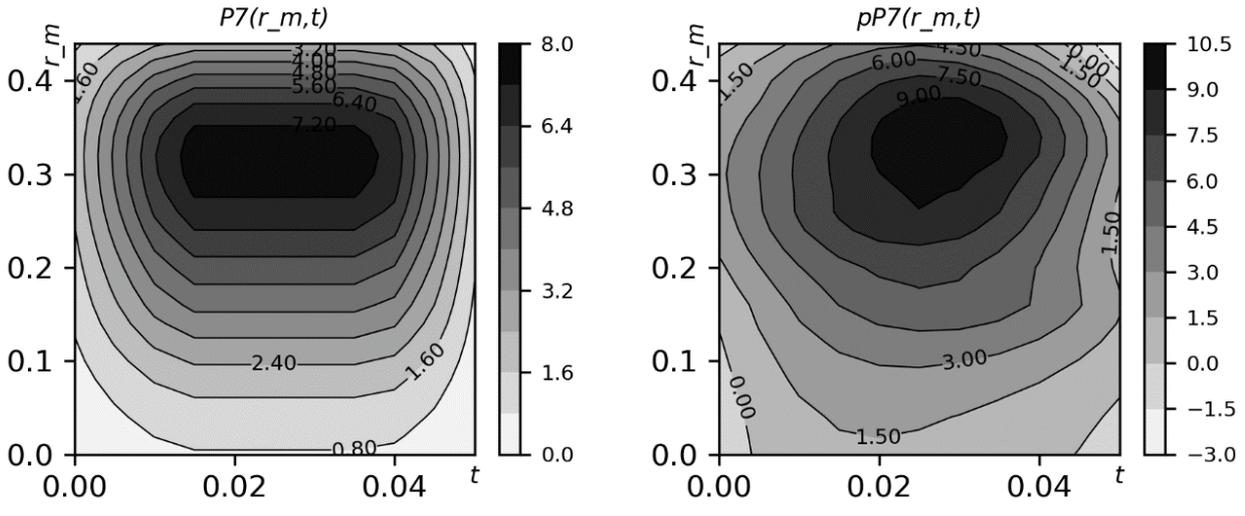

Fig. 1(d). Comparison of the left- and right-hand sides of Eq. (10).

For shot #53488, time interval from 7.5 s to 7.55 s is taken. The solution of the inverse problem gives $\bar{q} = 7$ ($f \sim 0.1$), and relative MSD = 74.1%. The results for sink and transmission functions are shown in Figure 2.

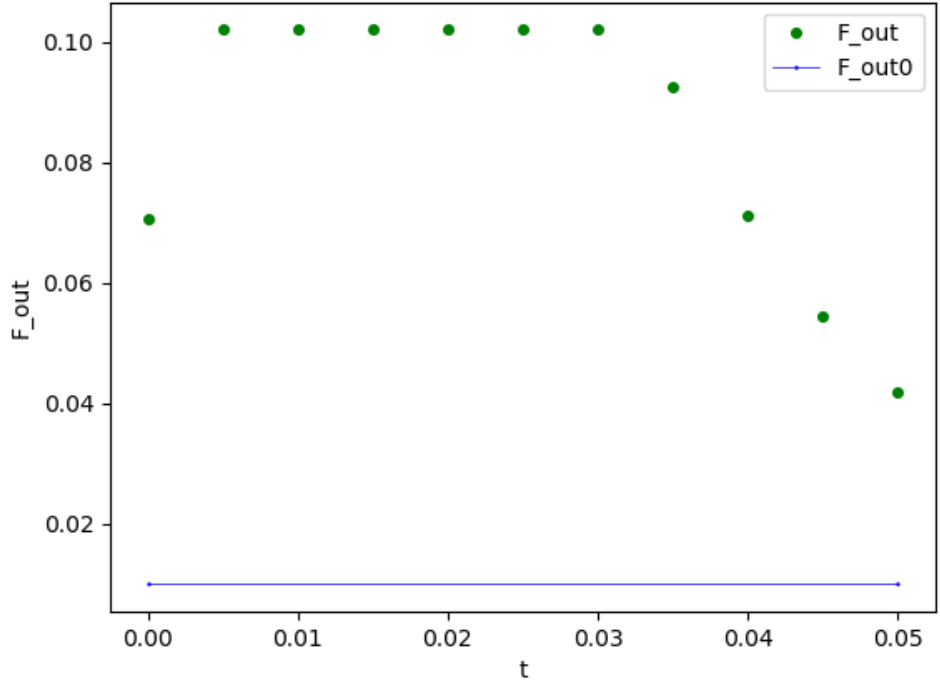

Fig. 2(a). Time dependence of transmission coefficient $F_{out} = 2\pi(1-R)$.

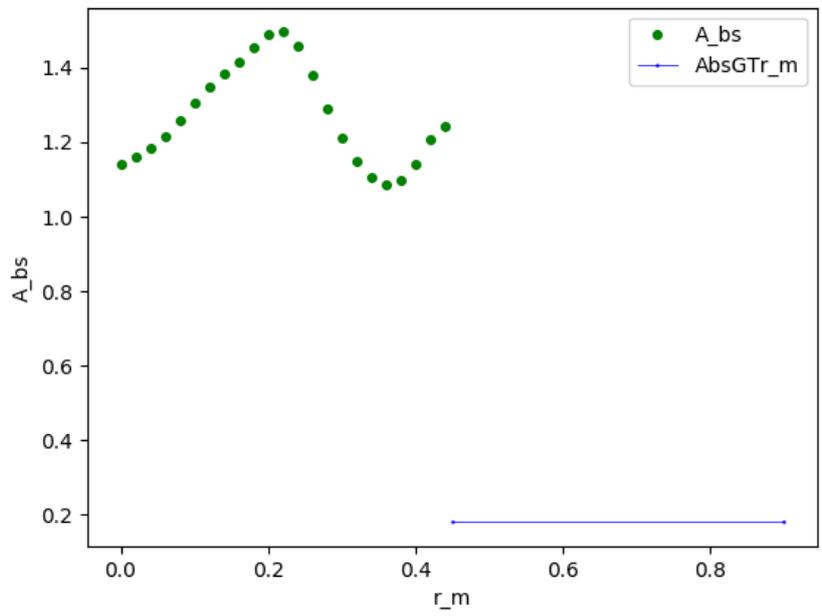

Fig. 2(b). The profile of absorption power as a function of the effective minor radius coordinate in the core plasma. The volume-average value in the residual plasma is shown.

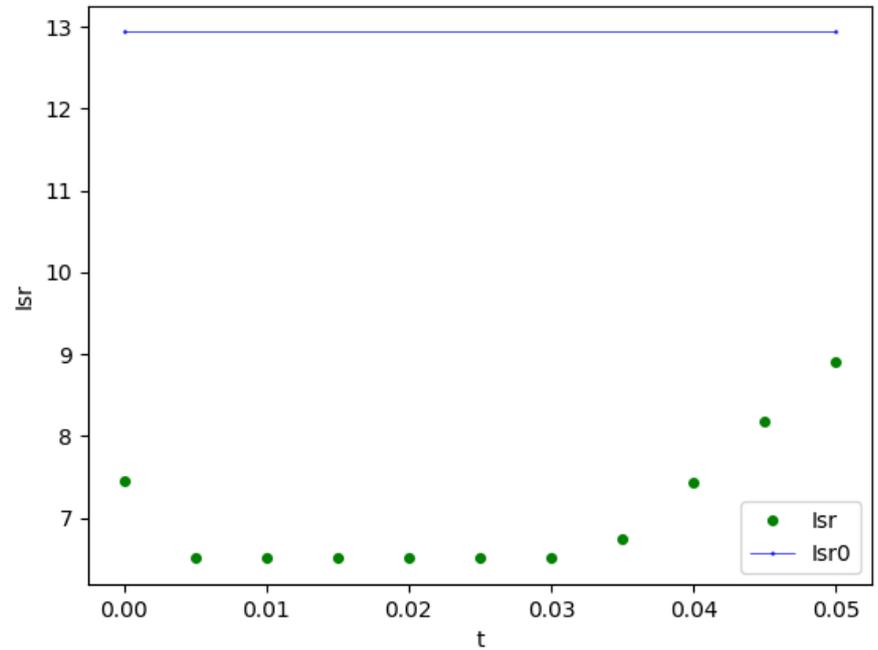

Fig. 2(c). Time dependence of the square root of normalized intensity, Eq. (4).

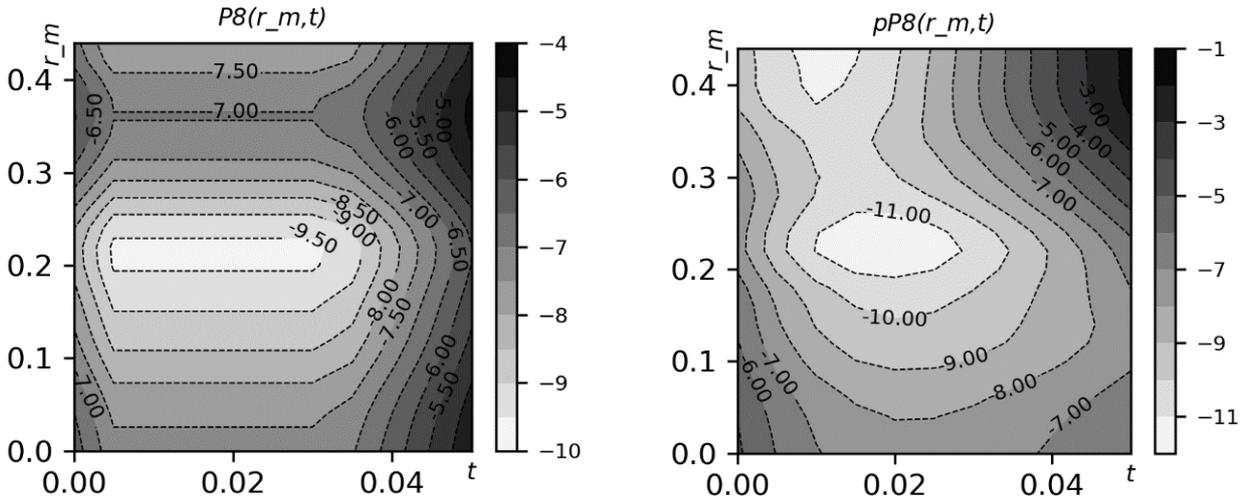

Fig. 2(d). Comparison of the left- and right-hand sides of Eq. (10).

### 4.2. Tokamak TFTR

Here we apply the model (1)-(10) to the data of the Ohmic discharge #88076, taking $t_0 = 37\ s$, $t_{min} = 3.72\ s$, $t_{max} = 3.74\ s$. Space-time dynamics of electron temperature in this discharge is presented in figures 2 and 3 in [4]. The solution of the inverse problem gives $\bar{q} = 3$ ($f \sim 0.15$), and relative MSD = 58.3%. The results for sink and transmission functions are shown in Figure 3.

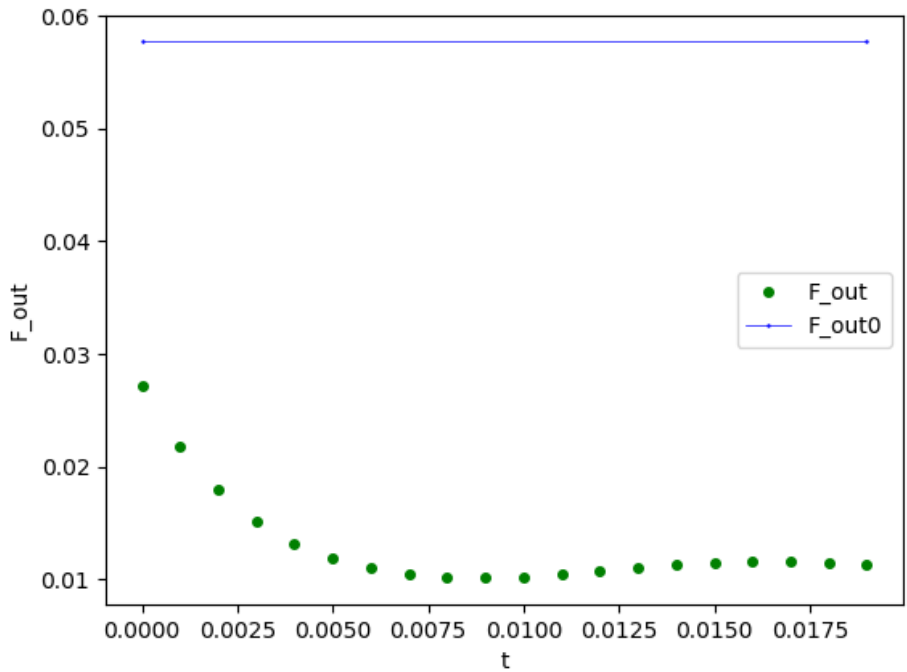

Fig. 3(a). Time dependence of transmission coefficient $F_{out} = 2\pi(1-R)$.

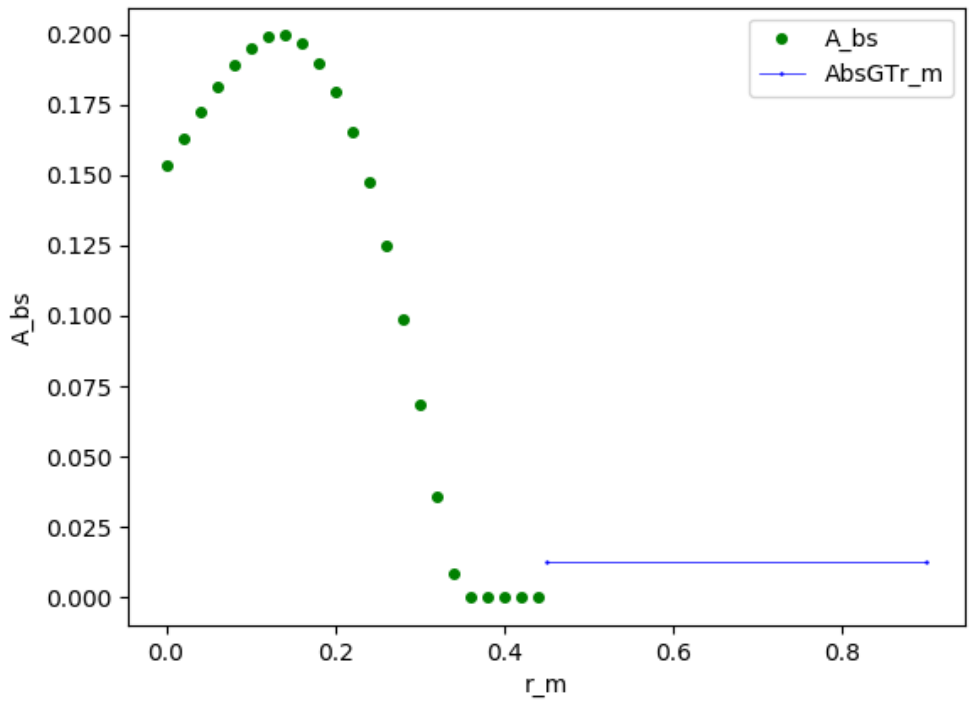

Fig. 3(b). The profile of absorption power as a function of the effective minor radius coordinate in the core plasma. The volume-average value in the residual plasma is shown.

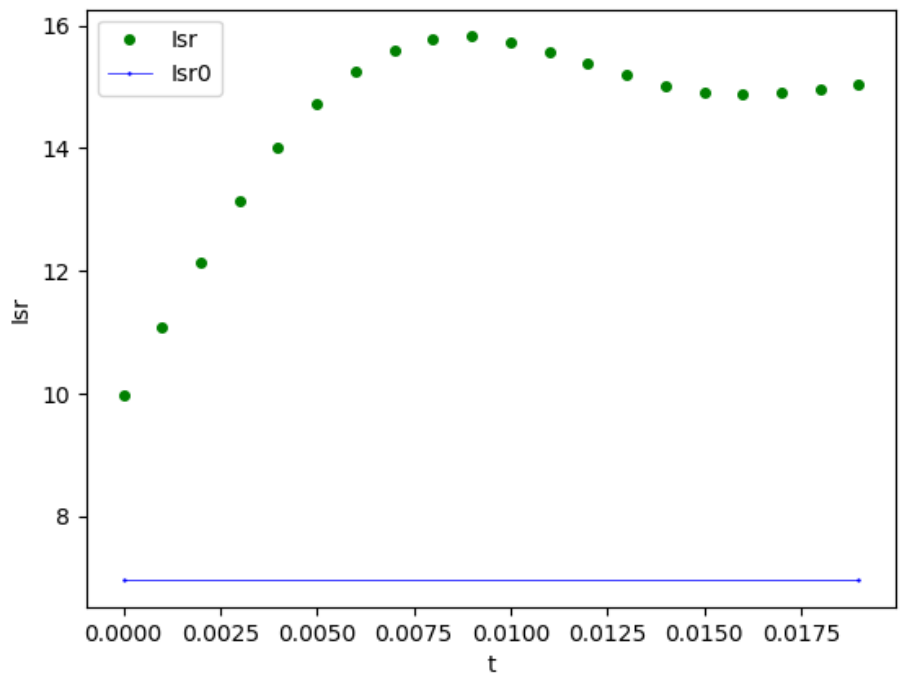

Fig. 3(c). Time dependence of the square root of normalized intensity, Eq. (4).

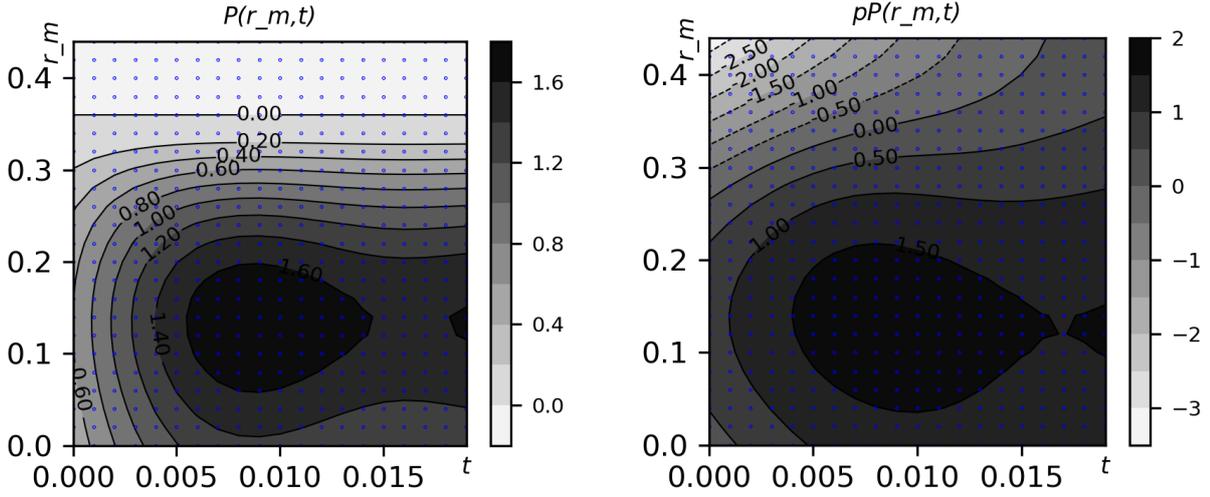

Fig. 3(d). Comparison of the left- and right-hand sides of Eq. (10).

### 4.3. Stellarator LHD

Here we apply the model (1)-(10) to the data of the discharge # 49708, taking $t_0$ = 2.797 s, $t_{min}$ = 2.8 s, $t_{max}$ = 2.815 s. Space-time dynamics of electron temperature and density in this shot with auxiliary plasma heating (2 MW power from injection of a neutral beam and 1 MW from injection of electron cyclotron waves) is presented in Figure 1 in [13] and Figure 1 in [14]. The solution of the inverse problem gives $\bar{q}$ = 21 ($f \sim 0.2$), and MSD = 10.6%. The results for sink and transmission functions are shown in Figure 4.

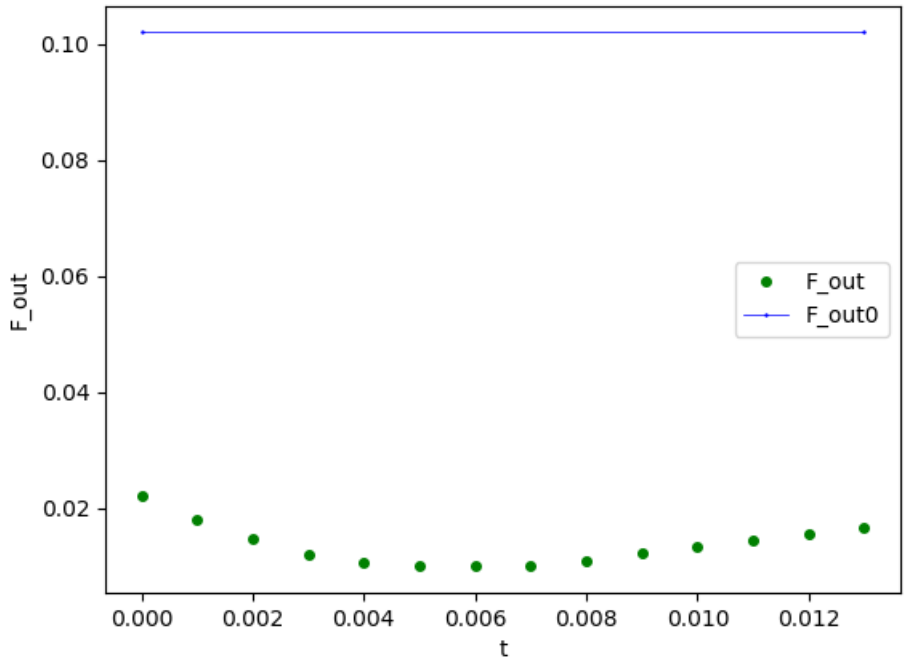

Fig. 4(a). Time dependence of transmission coefficient $F_{out} = 2\pi(1-R)$.

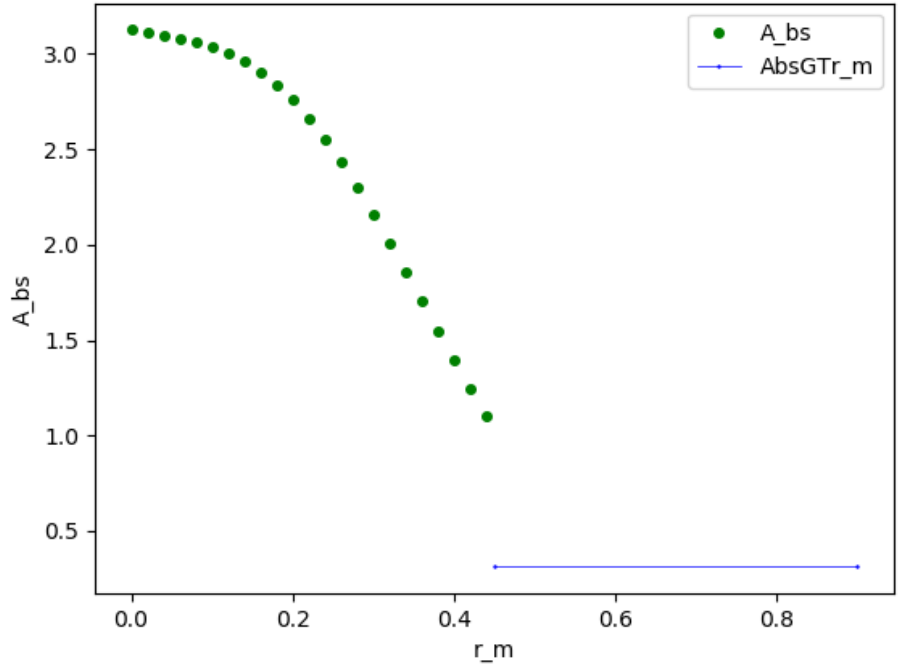

Fig. 4(b). The profile of absorption power as a function of the effective minor radius coordinate in the core plasma. The volume-average value in the residual plasma is shown.

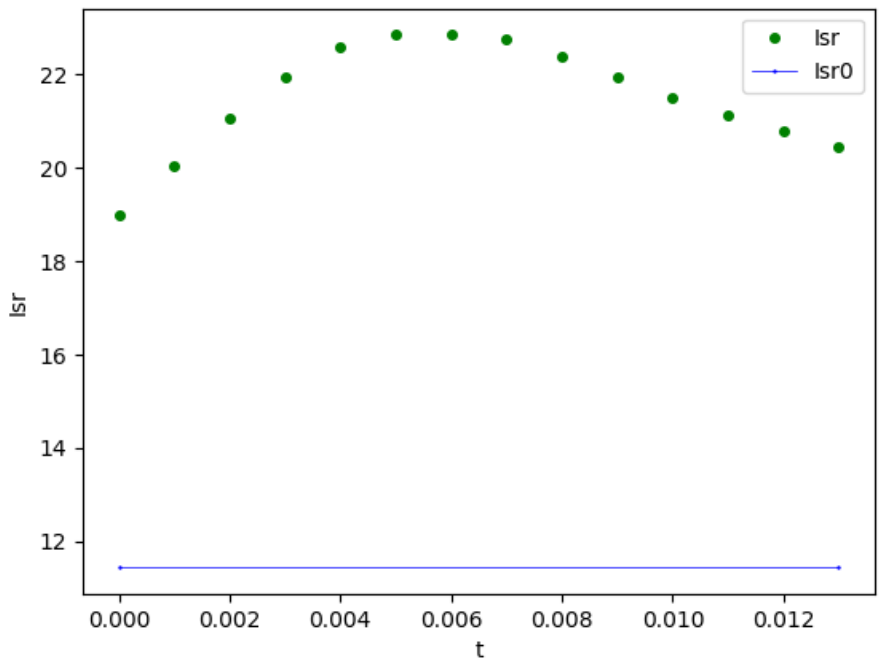

Fig. 4(c). Time dependence of the square root of normalized intensity, Eq. (4).

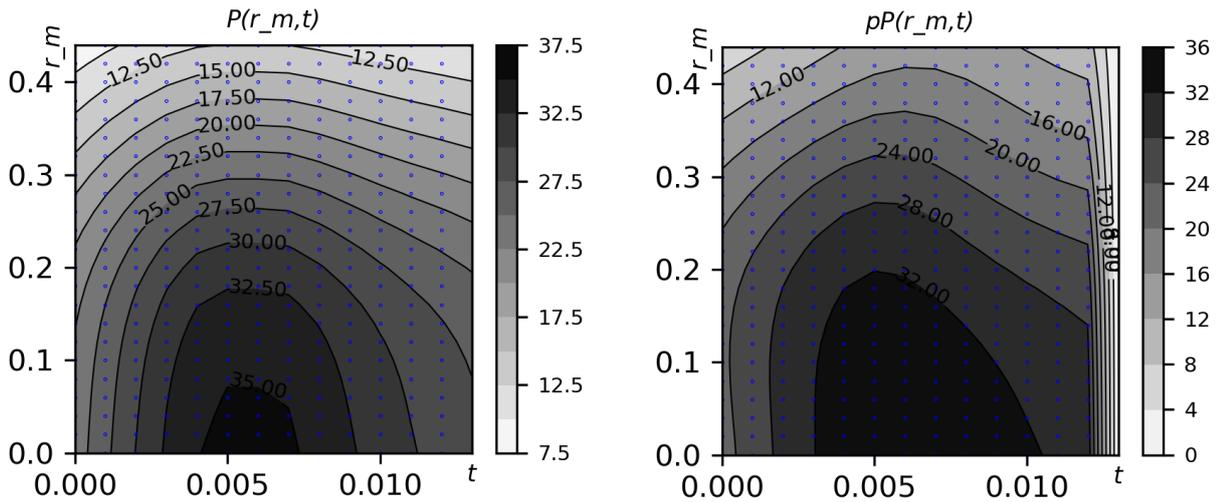

Fig. 4(d). Comparison of the left- and right-hand sides of Eq. (10).

## 5. Conclusions

The model (1)-(10) modifies -- in the points A, B, and C of Sec. 2 -- the model [11, 12] of nonlocal (superdiffusive) energy transfer by emission and absorption of electromagnetic (EM) waves with a long free path and strong reflection of waves from the walls of vacuum chamber. The model (1)-(10) assumes high internal reflections of waves and is compatible with the "wild cable" transport of TEM waves along magnetically-bound skeletal nanostructures [20, 21]. An inverse problem for recovery of the source and sink of waves, and internal reflectivity, is formulated and solved. Preliminary results of analyzing the data from tokamaks JET [19] and TFTR [4, 15], and stellarator LHD [13, 14] show that the model (1)-(10) may qualitatively explain the evolution of electron temperature in the core plasma during the initial stage of fast nonlocal transport events of the "cold pulse" type, because the model is compatible with a strong time variation of the reflectivity of the waves, and with high values of reflectivity as compared to those of the vacuum chamber wall.

## Acknowledgements

This research was partially funded by Russian Foundation for Basic Research (RFBR), grant numbers 18-07-01269-a, 18-07-01175-a.

The authors are grateful to A.S. Tarasov and V.V. Voloshinov for the help in using the Everest platform and the computing resources of the Center for Distributed Computing (http://distcomp.ru) of the Institute for Information Transmission Problems (Kharkevich Institute) of Russian Academy of Science, V.S. Neverov, for helpful discussions on data analysis, and A.P. Afanasiev, for the support of collaboration between the NRC "Kurchatov Institute" and the Center mentioned above.